\documentclass[11pt,tightenlines,eqsecnum,floats,aps,nofootinbib,prd,showpacs,twocolumn]{revtex4}

\usepackage{amsmath,amssymb,amsfonts,amsthm,amscd}
\usepackage{graphicx}
\usepackage{enumerate}
\usepackage{colordvi}
\usepackage{units}
\usepackage{epsfig}
\usepackage{natbib}
\usepackage{enumerate}
\usepackage{colordvi}

\def\be{\begin{equation}}
\def\ee{\end{equation}}
\def\ba{\begin{eqnarray}}
\def\ea{\end{eqnarray}}

\def\L{\mathcal{L}}
\def\f{\frac}
\def\k{\kappa}

\def\H{\mathcal{H}}

\begin{document}

\title{Loop Quantum Cosmology and the Fine Structure Constant}
\author{David Sloan$^1$}
\email{djs228@hermes.cam.ac.uk}
\affiliation{$^{1}$ DAMTP, Centre for Mathematical Sciences, Wilberforce Rd., Cambridge
University, Cambridge CB3 0WA, UK}

\begin{abstract}
The cosmological implications of introducing a variation to the fine structure `constant', $\alpha$ are examined within the context of Loop Quantum Cosmology. The evolution of $\alpha$ is described using the model introduced by Bekenstein, Sandvik, Barrow and Magueijo, in which a ghost scalar field produces the variation. The dynamics of the system are examined in flat and closed cosmological settings. Matter consisting of the scalar field and radiation are examined with a thermodynamically motivated coupling between the two, which can lead to a series of bounces induced by both the negative density effects of the ghost field and the loop effects. 
\end{abstract}

\maketitle

\section{Introduction}

The concept of a ``phoenix'' universe in which the big bang is replaced by a bridge between contracting and expanding phases of the cosmos goes back Tolman \cite{Tolman} and Lema\^itre \cite{Lemaitre}. Such models remove the initial singularity, replacing it by some causal link between two branches of solutions to the Friedmann equation. Since the powerful singularity theorems of Hawking and Penrose \cite{Hawking:1969sw} show geodesic incompleteness exists under very general circumstances, it is clear that these corrections can only occur if significant departures are made from General Relativity. There are many ways in which these can be brought about, but broadly speaking most fall into one of two camps: Corrections to Einstein's equations, usually as a result of some underlying quantum theory of gravity, or exotic matter which breaks the energy conditions invoked by the singularity theorems. In this paper the conjunction of two such models will be examined - the induced effective matter field arising from a model of variation of the fine structure constant, and Loop Quantum Cosmology.

The promotion of the fine structure constant, $\alpha$ to a field on space-time was investigated by Bekenstein \cite{Bekenstein:1982eu}. This was achieved by extending Maxwell's equations to include an electron charge which is allowed to vary between space-time points. Sandvik et at \cite{Sandvik:2001rv} expanded upon this model by adding gravity, and this resulted in the Bekenstein-Sandvik-Barrow-Magueijo (BSBM) cosmological models which are the counterparts to the Jordan-Brans-Dicke models of varying Newton's constant, $G$ \cite{Brans:1961sx}. 

The BSBM models have enjoyed success in their compatibility with high redshift quasar spectra \cite{Webb:2010hc} which provide data on historical variations in $\alpha$. Further achievements of the theory have included the description of an effective scalar field model describing the variations in $\alpha$ which is coupled to other dynamical matter fields \cite{Barrow:2011kr}, and an account in which this field is given a potential \cite{Barrow:2008ju}. The effect of the variation of $\alpha$ is to introduce a negative `ghost' energy density, the result of which is that big bang like singularities are resolved. Barrow, Kimberly and Magueijo showed in \cite{Barrow:2004ad} that this can lead to stable oscillations about an Einstein-static universe in the case of a closed FRW model, and this was extended to include locally rotationally symmetric Bianchi IX models in \cite{Barrow:2013qfa}, in which it was shown the field couplings introduce a natural isotropization on dynamics and lead to cosmological inflation.   

The resolution of singularities due to the inclusion of ghost fields is not a new - an extensive analysis of such models is given in \cite{Gibbons:2003yj}. Furthermore, it is well known that such fields are not well suited to quantization, resulting in negative norm states. However, the ghost field induced by the BSBM models is the result of an effective description of underlying dynamics, and as such we shall take the view that they should not be quantized. 

The Loop Quantum Cosmology (LQC) program is an extraction of cosmological dynamics from the underlying theory of Loop Quantum Gravity \cite{APS,APSV,Ashtekar:2011ni}. Since its inception, LQC has enjoyed many successes: The theory has been shown to be free from strong singularities \cite{ParamLQCNeverSingular,ParamFrancesca,ParamB1Singular,Corichi:2012xr}, with the big bang being replaced by a quantum bridge between expanding and contracting classical trajectories, dubbed the `big bounce'. The quantum theory admits an effective, semi-classical description \cite{Taveras} which closely matches the full dynamics even when quantum effects dominate. Outside the isotropic models, Bianchi geometries have been examined extensively in LQC, both within the quantum framework \cite{EdB1,EdB2,EdB9} and the semi-classical effective framework \cite{AlexB1,DahWei,Cailleteau,DahWei2,Maartens,Gupt,AlexB2,Gupt2,AlexB9}.  

The corrections to Einstein's equations induced by loop quantization are typically on the order of the Planck scale, with cosmological implications suppressed by a factor of the Planck density. It is therefore a difficult task to obtain observable consequences of the theory which deviate significantly from those of General Relativity. Significant progress to this end has been made in the context of inflation, which has been shown to be highly probably within the theory \cite{Measure,Corichi:2010zp,Measure2} and capable of producing a perturbation spectra compatible with the Cosmic Microwave Background \cite{AbhayWillIvan}. 

In this paper we will examine the evolution of $\alpha$ in LQC. The paper is laid out as follows: In Sec. \ref{Setup} we outline the geometry and matter models under consideration, and the corrections to the Friedmann equation resulting from loop quantization. In Sec. \ref{Flat} we establish the dynamics in the case of a flat ($k=0$) geometry and derive conditions differentiating bounces induced by matter and geometric effects.  Sec. \ref{Closed} deals with closed ($k=1$) geometries which undergo a sequence of bounces and recollapse and Sec. \ref{Discussion} gives some concluding remarks. 

\section{Setup}
\label{Setup}

In this paper we will consider two large deviations from the usual dynamics of General Relativity.  We consider a Roberston-Walker space-time, where the dynamics is given by the effective equations of LQC coupled to standard matter fields and a ghost scalar field which plays the role of the varying fine structure constant as per BSBM models. In the following we set out the dynamics resultant from each of these departures. 

\subsection{Geometry}

We consider a Hamiltonian formulation of cosmology in which our basic variables consist of the volume of a fiducial cell, $v$ and its conjugate momentum. \footnote{Within GR the conjugate momentum to the volume is the Hubble parameter, however in LQC this relationship is broken, as is shown below.} We will examine the flat ($k=0$) case initially as here it is straightforward to establish the roles that various fields play in dynamics, then extend to the closed ($k=1$) system. Let us fix $\k=8 \pi G$ as is conventional.

Our fundamental Poisson brackets are given by

\be \{v,b\} = \k \gamma/2 \ee

In this $\gamma$ is the Immirzi parameter. The value of $\gamma$  does not play a role in the dynamics of GR, but is relevant to the quantum theory. The Hamiltonians we consider are:

\be \H= \H_g^k + \H_m \ee

in which $\H_g^k$ is the gravitational Hamiltonian for curvature $k$ and $\H_m = v \rho$ the matter content. 

In the flat case \cite{APS}

\be \H_g^0 = -\f{3}{\k} \f{v sin^2(\lambda b)}{\gamma^2 \lambda^2} \label{Hflat} \ee

In the closed case \cite{APSV,Corichi:2011pg,ParamFrancesca} \footnote{In \cite{Corichi:2011pg} there are two separate definitions of the Hamiltonian describing the effective dynamics of a loop quantized closed cosmology, corresponding differences arising from basing the quantization upon holonomies or connections. In this paper we follow the holonomy based quantization, and note that although there will be minor quantitative differences in the behaviour of the fine structure constant depending on this choice, the qualitative features induced by the loop bounces will be the same} :

\ba \H_g^1 = &-&\f{3}{\k} \f{v}{\gamma^2 \lambda^2} \big(sin^2(\lambda b' - D) \nonumber \\
             &-& sin^2 (D) + (1+\gamma^2)D\big) \label{Hcurv}  \ea

In which $D= \f{\lambda \sigma}{v^{1/3}}$, and $\sigma=(2 \pi^2)^{1/3}$. We make the simple transformation $b=b'-D/\lambda$ to simplify the system, noting that this has no effect on our Poisson bracket: 

\ba \H_g^1 = -\f{3}{\k} \f{v}{\gamma^2 \lambda^2} \big(&sin^2(\lambda b)& - sin^2 (\f{\lambda \sigma}{v^{1/3}}) \nonumber \\  + &(1+\gamma^2)&\f{\lambda \sigma}{v^{1/3}} \big) \ea

From the Hamiltonian constraint we find our equations of motion:

\be \dot{v} = \f{3 v}{2 \gamma \lambda} sin(2 \lambda b) \ee

which is independent of the curvature, and $\dot{b}$ being

\ba \label{bdv} \dot{b}= &-&\k \gamma (\rho+P)/2 \nonumber \\
                \dot{b}= &-&\f{\sigma D}{2 \gamma \lambda^2} sin(2D) \nonumber \\
                         &+& (1+\gamma^2) D^2 -\k \gamma (\rho+P)/2  \ea

in the flat and closed cases respectively.
Here we have defined the pressure $P$ in the usual way, $P=\f{\partial H_m}{\partial v}$. In order to make a more direct comparison with the existing literature on cosmology, let us write our equations of motion in terms of the scale factor $a=\f{v^{1/3}}{\lambda \sigma}$ into which we have absorbed constants to clean up the algebra. Thus we obtain 

\be H = \f{\dot{a}}{a} = \f{1}{2 \gamma \lambda} sin (2 \lambda b) \ee
\be \dot{b} = -\f{1}{2 \gamma \lambda a} sin (\f{2}{a}) + \f{1+\gamma^2}{a^2} - \k \gamma (\rho+P)/2 \ee
 in the flat case, $\dot{b}$ is as in equation \ref{bdv}.

\subsection{Matter}

We consider matter which consists of a radiation field, and as per the BSBM model, a scalar dielectric field. As will be shown below, the net effect of the scalar dielectric field will be to act as an effective ghost scalar field.

\subsubsection{The BSBM Model} 

We take the fine strucutre constant $\alpha$ to be dynamical. It has been claimed that there is evidence of spatial dipole in $\alpha$ \cite{Webb:2010hc}, therefore within the context of relativity this can be naturally extended to a variation in space-time. The BSBM model differs from most approaches to varying `constants' through the introduction of the scalar field. Past approaches often made $\alpha$ dynamical ``by hand'', simply setting $\alpha=\alpha(x)$ say, in equations where $\alpha$ was previously constant. These suffered from a lack of internal consistency between action and dynamics, and broke conservation laws. 

In the BSBM model the effect of varying $\alpha$ is introuced through an evolving scalar field, $\psi$, which alters the charge of the electron via

\be e = e_0 e^{\psi} \ee  

in which $e_0$ is the (dimensionful) charge at some fixed time. Since we will be dealing with homogeneous, cosmologies, we allow $\psi$ to be a field which varies with time but is uniform in space. 

The action for our matter fields is given

\be S = \int \sqrt{-g} \big( \L_m + \L_\psi + e^{-2 \psi} \L_{em} \big) \ee

In this $\L_\psi = - \f{\omega}{2} \partial_u \psi \partial^u \psi$ determines the evolution of the scalar field. $\L_{em}$ is the usual electromagnetic action $-\f{1}{4} f_{\mu \nu} f^{\mu \nu}$  and $\L_m$ some standard matter Lagrangian unrelated to the fine structure constant. 

Thus in the evolution of matter density, $\psi$ enters the Friedmann equation in the form of a scalar field with only a kinetic term (ie a stiff fluid). The sign of $\rho_m \omega/ \L_m$ determines the direction of time evolution of $\alpha$ during the cold dark matter dominated phase of the universe, where $\alpha$ increases for negative sign, and decreases for positive. Fitting to the Keck data \cite{Sandvik:2001rv} sets $\omega$ to be negative, and thus $\psi$ takes the form of a ghost field, a stiff fluid with negative energy density. Henceforth it will be convenient to consider the effective field describing this behaviour, being $\rho_\psi$, where

\be \label{psiEOM} \rho_\psi =  \omega \dot{\psi}^2 \ee 

\subsubsection{Matter Couplings}
The matter content will be taken to be the sum of perfect fluids each with energy density $\rho_i$ and pressure $P_i = w_i \rho_i$ (no summation) with dynamics goverened by 

\be \dot{\rho_i} + 3 H (\rho_i + P_i) = \Delta_i \label{MatterEoM} \ee

Here we model couplings between the fluids by the terms $\Delta_i$ and the sum over all fluids of these terms is zero, so that in energy transfers between fluids total density is conserved. When there is no coupling between matter fields these energy densities have their usual dependences upon scale factor: 

\be \rho_i = \f{\Gamma_i}{a^{3(1+w_i)}} \ee

in which the $\Gamma_i$ are constants.

Throughout this paper we will take the energy density to consist of a (ghost) scalar field given by the BSBM Lagrangian \cite{Barrow:2004ad} representing the varying fine structure constant and radiation. Thus our energy density is the sum:

\be \rho = \rho_\psi+\rho_r \ee

Since the scalar field is a ghost field, i.e.  $\rho_\psi$ is negative, the total energy density will vanish at finite scale factor, and with non-vanishing net pressure. Thus we see that all the usual energy conditions are violated for the net density, which in the case of GR is how the singularity theorems are evaded. \footnote{It is a simple calculation to show that if matter sources satisfy, e.g. the weak energy condition the their sum will satisfy the condition also, \textit{provided the density of each source has the same sign}. Once a ghost field is introduced this fails to hold in general, and will be violated at the transitions between dominance of one field and another of opposite sign.}

\subsubsection{Curvature}

 The treatment of curvature within the effective equations of LQC differs from its role in the classical Friedmann equations. In a classical system, the Hamiltonian for a closed geometry is given:

\be \H_{cl} = -\f{3}{\kappa \gamma^2 } vb^2 - \f{3}{\kappa} v^{1/3} + v \H_m \ee

Here it is apparent that one could capture the effects of curvature by absorbing the second term in this equation into the matter Hamiltonian, $\H_m$ resulting in an effective fluid with energy density given by

\be \rho_k = \f{\Gamma_k}{a^2} \ee

in which $\Gamma_k = -3/\kappa$, and hence pressure $P_k = -\rho_k/3$. However, within the effective LQC equations, if curvature is modelled as a fluid, its equation of state varies over time. To see this consider defining $\H_k$ such that $\H_1=\H_0+\H_k = \H_0+v\rho_k$. Thus we find the effective energy density: 

\be \rho_k = \f{3}{\kappa \gamma^2 \lambda^2} \big( sin^2(\f{1}{a}) - \f{(1+\gamma^2)}{a^2} \big) \ee

and pressure 

\be  P_k = \f{3}{\kappa \gamma^2 \lambda^2} \big( \f{1+\gamma^2}{3a^2} + sin^2(\f{1}{a}) -\f{sin(\f{2}{a})}{a} \big) \ee

Thus the effective fluid has an equation of state $P_k = w_k \rho_k$ in which $w_k$ is a function of both scale factor $a$ and the value of the Immirzi parameter $\gamma$. For large values of either of these $w_k$ asymptotes to the classical value of $-1/3$, and always lies between $-2/3$ and $-1/4$. 

\section{Flat Geometries}
\label{Flat}

As a warm-up problem, let us consider the flat ($k=0$) geometries. The dynamics are determined by the modified Friedmann equation \cite{Taveras}, derived from \ref{Hflat}. We define the critical density $\rho_c = \f{3}{\k \gamma^2 \lambda^2} \approx 0.41 \rho_{planck}$ to be the maximum energy density accessible in LQC \cite{Ashtekar:2011ni}. 

\be H^2 = \f{\k \rho}{3} (1-\f{\rho}{\rho_c}) \ee

Note that the Hamiltonian constraint \ref{Hflat} ensures that $\rho \leq \rho_c$, so $H$ is real on all solutions. Let us begin by considering the matter content to consist of the effective scalar field of the BSBM model and radiation and leave these uncoupled. Thus our energy density obeys

\be \rho = \f{\Gamma_\psi}{a^6} + \f{\Gamma_r}{a^4} \ee

The flat Hamiltonian, $\H^0$ has a symmetry under a rescaling of $a$. In a flat universe the scale factor alone holds no meaning. It is measured with respect to a fiducial cell, the size of which should have no influence on physics. This is manifest in our system in the form of a symmetry between solutions under which $\{a,\Gamma_i\} \rightarrow \{\lambda a, \lambda^{1+w_i} \Gamma_i\} $, which keep physical quantites such as curvature, energy density and pressure fixed. Therefore, in specifying a solution in which $n$ fluids are present, only $(n-1)$ parameters are required. In the system considered here, there are only two fluids, hence their ratio $R=\Gamma_\psi / \Gamma_r^{3/2}$ uniquely determines a physical trajectory, and is invariant under the action of the symmetry.

\subsection{Bounces}

Since the scalar field is a ghost field, $\Gamma_\psi$ is negative. It is clear from our equations that there are two conditions for a bounce, $\rho=0$ and $\rho=\rho_c$, both of which can be achieved with non-zero matter content. Let us call these a `ghost' bounce if $\rho=0$ and a `loop' bounce if $\rho=\rho_c$. A ghost bounce occurs if on a trajectory $a=a_g=\sqrt{-\Gamma_\psi/\Gamma_r}$ is achieved. Maximizing the energy density across scale factor, we see that the maximum possible density can occur at $a=a_m = \sqrt{\f{-3 \Gamma s}{2 \Gamma r}} = \sqrt{3/2} a_g$. Here the energy density is $\rho_m = \f{4 \Gamma_r^3}{27 \Gamma_\psi^2}=\f{4}{27}R^{-2}$. Since $a_m$ is larger than the scale factor for a ghost bounce ($a_g$), it is apparent that if $\rho_m$ exceeds $\rho_c$ for a large collapsing universe there will be a loop bounce, otherwise there will be a ghost bounce. 

\subsection{Solutions}

\begin{figure}[ht]
\includegraphics[width=0.49\textwidth]{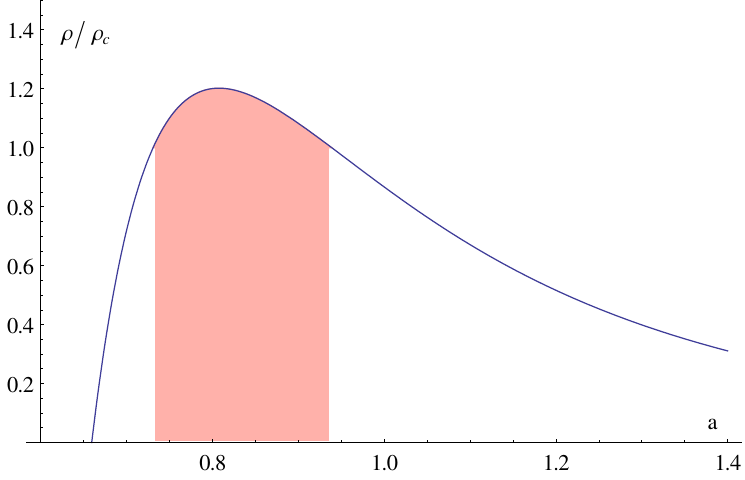}
\caption{Energy density profile as a function of scale factor, with the region forbidden by the constraint shaded. Here we see that solutions can exist in either of the two separate regions. For this graph, $\Gamma_\psi=-1$,$\Gamma_r=2.3$}
\label{regions}
\end{figure}

The range of $a$ visited by a dynamical trajectory is determined entirely by the profile of the energy density, with those regions in which $0 \leq \rho \leq \rho_c$ allowed by the constraint. If $\rho_m$ is smaller than $\rho_c$ this constitutes a single connected region. However, if $\rho_m$ is larger than $\rho_c$ the allowable region splits into two, as shown in figure \ref{regions}. 

Solutions remain trapped in a single region thoughout their evolution. In the left region the energy density increases with scale factor, and upon reaching the critical density the sytem undergoes a recollapse which is brought about by the loop corrections. These solutions are highly non-classical, as their behaviour at the largest scales is dominated by quantum effects. In the right region solutions are bounded from below by the forbidden region. Trajectories undergo a unique bounce at $\rho=\rho_c$. At low energy densities the loop effects become negligable, and solutions follow their classical counterparts.  

When field couplings are applied, the resulting dynamics are changed. The energy density profile shown in figure \ref{regions} is no longer fixed, as energy is transferred between the two fields. Let us consider the effect of such a transfer on a trajectory within the trapped region. Since the trajectory is bounded, its scale factor is always below that of the maximum energy density, hence $a < \sqrt{\f{-3 \Gamma_\psi}{2 \Gamma_r}}$. Let us denote the fluid parameters before the tranfer $\Gamma_\psi, \Gamma_r$ and after $\Gamma'_s, \Gamma'_r$.  Conservation of the energy density yields 

\ba \Gamma'_s &=& \Gamma_\psi- \Delta a^6 \\
    \Gamma'_r &=& \Gamma_r+ \Delta a^4 \ea

wherein $\Delta$ represents the amount of energy density transferred, which we shall assume to be small compared to the total energy density. Hence we find the change in $R$, to first order in $\Delta$:

\be \f{R'}{R} =1-\f{a^2}{2} \big( \f{2\Delta a^2}{\Gamma_\psi} + \f{3 \Delta}{\Gamma_r} \big) \ee 

\begin{figure}[ht]
\includegraphics[width=0.49\textwidth]{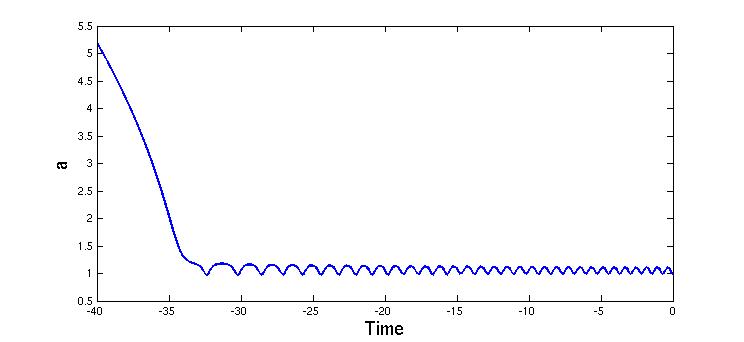}
\caption{A collapsing branch becomes trapped due to the transfer of energy density between fluids. Initial values $a=5.2$, $\lambda b=-0.09$, $\Gamma_\psi=-0.3$, $\Gamma_r=243$,$\Delta=0.01 \rho_\psi$}
\label{trap}
\end{figure}

Since $a < \sqrt\f{-3 \Gamma_\psi}{2 \Gamma_r}$ this terms is larger than unity for negative $\Delta$, and hence $\rho_m$ increases. On the other side of the excluded region, the scale factor will be larger than $\sqrt\f{-3 \Gamma_\psi}{2 \Gamma_r}$ and hence the maximum energy density will decrease. It is therefore possible for a collapsing solution to enter this region, since $\rho_m$ is initially lower than $\rho_c$, however whilst in the region the transfer of energy inceases $\rho_m$ to the point that the evolution becomes locked between ghost bounce and quantum recollapse. An example such trajectory is shown in \ref{trap}

\section{Closed Geometries}
\label{Closed}

Let us now consider the case of a closed Robertson-Walker geometry in which the dynamics are governed by Eq.\ref{Hcurv}. Following \cite{Corichi:2011pg} we can express the modified Friedmann equation in the $k=1$ case as 

\be H^2 = \f{\kappa}{3} (\rho-\rho_1)(1-\f{\rho-\rho_1}{\rho_c}) \label{CurvFried} \ee

in which $\rho_1 = \f{1+\gamma^2}{a^2} -sin^2 (\f{1}{a})$. Again, the Hamiltonian constraint \ref{Hcurv} ensures that $\rho-\rho_1 \leq \rho_c$ and thus $H$ is always real on solutions. 

\subsection{Bounce and Recollapse}

At a bounce or point of recollapse the Hubble parameter vanishes. From \ref{CurvFried} it is obvious that this is possible in two ways - the first when $\rho=\rho_1$ and the second when $\rho-\rho_1=\rho_c$. In the second case, it is clear from the equations of motion that $\lambda b=(n+1/2) \pi$ for some integer $n$, and corresponds to a loop bounce. In the prior case, the system can experiences either a ghost bounce or recollapse due to the energy density. Loop bounces or recollapses will occur if the second condition is ever met on a trajectory, otherwise at small scale factors the system will undergo a ghost bounce. The effective energy density profile, $\rho-\rho_1$ is similar to that in the flat case at small scales, and again the bound placed upon the density splits the scale factors allowed by the Hamiltonian constraint into two regions. In contrast to the flat case, the region containing large scale factors is here bounded, as it is classically. 

\subsection{Solutions}

\subsubsection{The Static Solution}

Similarly to the classical case \cite{Barrow:2004ad} there exists a static solution in which none of our dynamical variables evolve. Consider the case in which there is no coupling between our fields. This solution is given by $b=n\pi$ for integer $n$ and matter:

\ba \rho_\psi   = \f{3}{\kappa} &\big(&\f{2}{\gamma^2 \lambda^2} sin^2(\f{1}{a}) - \f{1+\gamma^2}{a^2} \nonumber \\
                                &-&\f{1}{2                      \gamma^2 \lambda^2 a} sin (\f{2}{a}) \big) \\
    \rho_r      = \f{3}{\kappa} &\big(&\f{2(1+\gamma^2)}{a^2} -\f{3}{\gamma^2 \lambda^2} sin^2(\f{1}{a}) \nonumber \\
                                &+&\f{1}{2\gamma^2 \lambda^2 a} sin(\f{2}{a}) \big)     \ea

for any scale factor $a$. The form of $\rho_\psi$ is such that it is always negative. Note that, although $b=n\pi$ would indicate that we are well into the classical regime in the flat case, here effects coming from quantum gravity are still significant for small scale factors. In comparison with the static solution of the classical system found by Barrow et al. in \cite{Barrow:2004ad}, these energy densities are larger by a factor of two for small scale factors, and reduce to become equal in the limit of large scale factor. \footnote{The definition of scale factor used here differs slightly from that of \cite{Barrow:2004ad}, and the two become equal upon translation of units.} This solution represents an energy density profile in which $\rho-\rho_1$ has only one root as a function of $a$, hence ghost bounce and recollapse are at the same point. Classically this corresponds to $\Gamma_i = -\Gamma_r^2/4$, but in terms the effective quantum description the profile involves trigonometric functions and so the condition cannot be solved analytically. 

In general the equations of motion are too complex to find exact dynamical solutions. However, one can perform a pertubative analysis about the static solution. Let us simplify the algebra here, setting $\kappa=\gamma=\lambda=1$ To first order one finds after an arduous calculation:

\be \ddot{\delta a} = \f{\delta a}{a^2} \big(-\f{9 a}{2} sin(\f{2}{a}) + cos(\f{2}{a}) + 12 a^2 sin^2(\f{1}{a}) - 8 \big) \label{daoscill} \ee

\begin{figure}[th]
\includegraphics[width=0.49\textwidth]{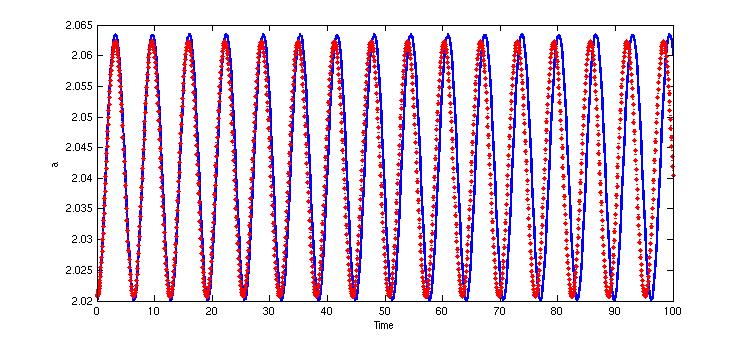}
\caption{Scale factor of system perturbed about static solution (solid, blue) and fit by sine using the frequency obtained from eq. \ref{daoscill} (dotted, red).
Initial values: $a=2.02$, $H=0$, $\rho_\psi=-0.384$, $\rho_r=0.781$, $\Delta=0$}
\label{Oscil}
\end{figure}

Here we have assumed that the perturbation leaves the $\Gamma_i$ fixed, but preserves the constraint. Hence we have an equation of the form $\ddot{\delta a} = - \omega^2 \delta a$ - a harmonic oscillator. An examination of the right hand side of eq. \ref{daoscill} shows that $\omega^2$ is indeed positive. A typical solution is shown in figure \ref{Oscil} in which the frequency for the fit used is that given by eq. \ref{daoscill}. Similar oscillatory behaviour can be found for the evolution of $\delta b$ and the matter degrees of freedom (all of which have the same frequency). 

\subsubsection{Evolving Solutions}

With no coupling between our fields, each solution is a cyclic trajectory undergoing recollapse at a maximum scale factor and either a loop or ghost bounce at minimum scale factor. When the fields are coupled in such a way that the second law of thermodynamics is obeyed, a system which undergoes loop bounces will eventually evolve into one which undergoes only ghost bounces. To see this, consider the effect of energy transfer in equation \ref{MatterEoM}. Energy density is transferred between fields in such a way that the total density is unaffected. Considered in terms of trajectories with no transfer, the net effect is to reduce the total energy density at scale factors lower than that at which the transfer occurs, and increase it in those above. For the simple, two field, system in consideration here, this reduces $\Gamma_\phi$ and increases $\Gamma_r$. Thus at the scale factor at which the previous bounce occured, the energy density is lower than previously, and hence the bounce must occur at smaller $a$.  However, as this value becomes progressively smaller, the effect of the ghost field becomes stronger at the minimum, since $\rho_\psi=\Gamma_\phi a^{-6}$ and $\Gamma_\phi$ has increased in magnitude (recall that $\Gamma_\phi<0$). Thus the maximum energy density $\rho_m$ across all possible values of the scale factor decreases, and eventually reaches the point at which it is no longer possible to satisfy $\rho=\rho_c-\rho_1$ for any $a$, and loop bounces stop, being replaced by ghost bounces. When the ghost bounces take over, the minimum scale factor rises, since the increase in $\Gamma_i$ and reduction of $\Gamma_r$ mean that $\rho=0$ at larger $a$. 

\begin{figure}[ht]
\includegraphics[width=0.49\textwidth]{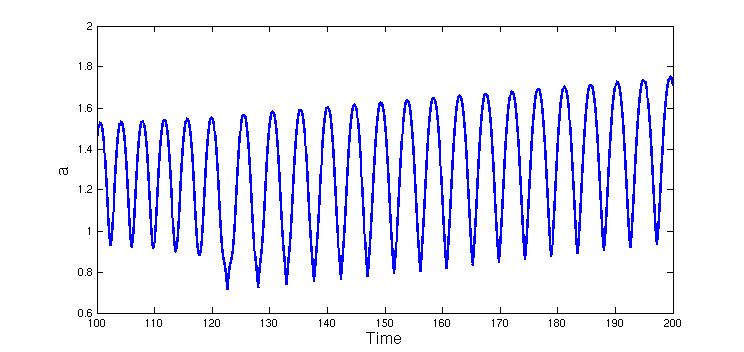}
\includegraphics[width=0.49\textwidth]{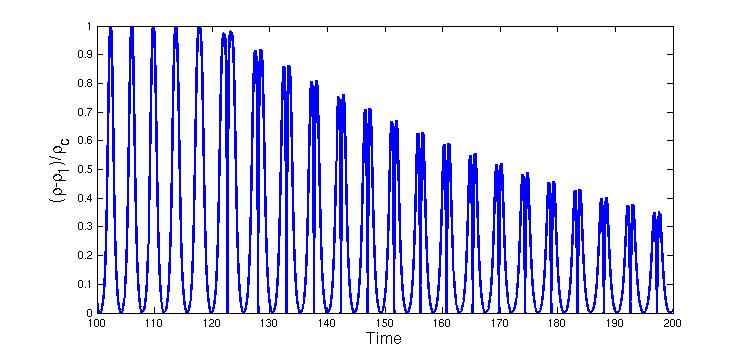}
\caption{Scale factor (above) and $\rho-\rho_1$ (below) against time. To begin with the system undergoes a series of loop bounces, but as energy is transferred the system begins undergoing ghost bounces. The scale factor at the bounce decreases throught the series of loop bounces, then increases during ghost bounces. Initial Values: $a=1$, $H=0$, $\rho_\psi=-0.562$, $\rho_r=4$, $\Delta=0.01 \rho_\psi$}
\label{Transfer}
\end{figure}

The behaviour of the fine structure constant throughout the evolution varies with the trajectory of the scale factor. During a ghost bounce there is a large change in $log(\alpha)$, however during a loop bounce the relative change is much smaller. This is apparent because the velocity of $\alpha$ is determined by $\rho_\psi$. For a fixed $\Gamma_\psi$, $\rho_\psi$ is maximized during a bounce, since at this point the scale factor is at its smallest. However, during the progression from a series of loop bounces to ghost bounces the field coupling is increasing the magnitude of $\Gamma_\psi$. Thus we find that the largest change happens during the series of ghost bounces. If one extends the solutions to the past, the dynamics become dominated by the radiation field, with the system experiencing an unending sequence of loop bounces and recollapses, and $\alpha$ becomes constant. Thus we find a significant correction to the past behaviour from using LQC: The evolution of $\alpha$ stops. In the classical case $\alpha$ continues to fall endlessly to the past, since each bounce is a ghost bounce, and thus $\rho_p$ is large there. The effect of the loop corrections is to remove this dominance at small scale factor, and so the change in $\alpha$ to the infinite past is a finite amount. 

\begin{figure}[ht]
\includegraphics[width=0.49\textwidth]{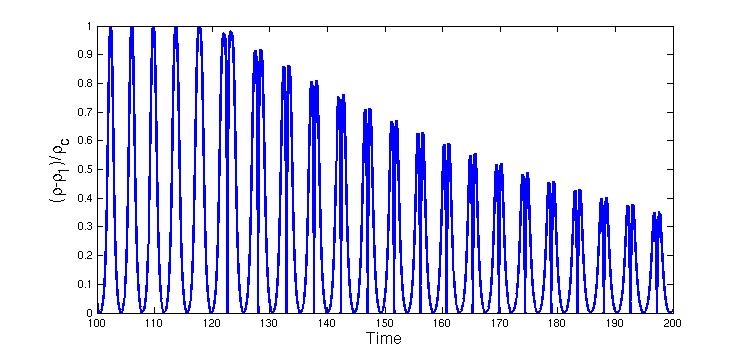}
\includegraphics[width=0.49\textwidth]{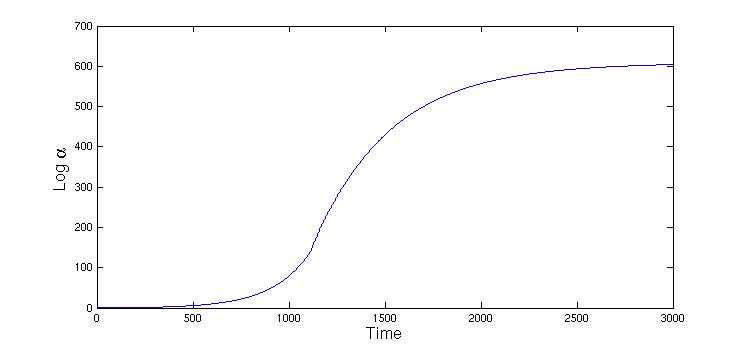}
\caption{Scale factor (above) and $Log (\alpha)$ (below) against time. A much longer evolution similar to that in the previous diagram. Again the system begins with a series of loop bounces and classical recollapses. We see that the fine structure constant does not change largely throughout the loop bounces, yet undergoes a large transition once the ghost bounces take over, eventually settling in a large universe. Initial Values: $a=1.02$, $H=0$, $\rho_\psi=-2 \times 10^{-5}$, $\rho_r=3.34$, $\Delta=0.01 \rho_\psi$}
\label{psivolve}
\end{figure}

To see this in more detail, consider the evolution of the ghost field, determined by equation \ref{MatterEoM} with $\Delta=s \rho_\psi$ for some fixed $\Delta$. Then, one can consider the evolution of this field across a complete cycle from one bounce to the next. The effective (time averaged) Hubble parameter here is zero, and so the main contribution to the evolution comes from $\Delta$, and hence $\rho \approx \rho_0 e^{st}$ for a constant of integration $\rho_0$. Thus from \ref{psiEOM} we find that $\psi=\psi_0 +\psi_1 e^{st/2}$ for some constants $\psi_0, \psi_1$, and we see that the value of $\alpha$ has undergone a finite change to the infinite past.  

\section{Discussion}
\label{Discussion}

Both LQC and the BSBM model of varying $\alpha$ give rise to bounces in the place of singularities. In the case of simple FRW models these bounces happen under different conditions - with loop bounces occuring when the energy density approaches the Planck scale, and ghost bounces occuring at zero net energy density. In this paper we have established that in the type of bounce which occurs is a function of the distribution of matter, giving a simple condition for its determination in the case of a flat geometry. 

Although it can be seen that the introduction of a ghost scalar field leads to singularity resolution in some cases, this does not remove the role of quantum gravity. Within the matter model under consideration the energy density and curvature can be arbitrarily large classically, so reaching into the Planck regime in which quantum gravity is expected to play a significant role in dynamics. It is therefore natural to examine the role played by LQC in models are classically non-singular as well as those which are singular. 

The conjunction of the two models leads to some unexpected behaviour: The loop effects can lead not only to bounces, but also to recollapse even within flat geometries. This effect, which becomes apparent on examining the allowed regions within Fig. \ref{regions} is not present when either model is considered separately, but rather depends on having both a maximum energy density (a loop effect) and the non-monotonicity of net energy density (a ghost effect). Field couplings can lead to large collapsing universes becoming trapped within this region. 

In the closed case we found significant departures from the GR behaviour. Again, the region in which quantum recollapse happens is present. As seen previous, cyclic solutions exist, and with field couplings these can evolve from being a sequence of small oscillations into a large universe. However, since loop effects lead to a bounce under certain matter distributions, which become dominant at early times, the evolution of $\alpha$ slows to the past, and thus a finite change happens across all time. This contrasts with the GR case in which $\alpha$ tends towards $-\infty$ to the past. Since the role of the effective ghost field is neither dominant during loop bounces nor at large scales, it is only during ghost bounces that large changes in $\alpha$ happen. This analysis relies crucially on understanding the dynamics of LQC, singularity resolution by other means will likely lead to different effects.
 
\section*{Acknowledgements}

This work was supported by a grant from the Templeton Foundation. The author is grateful to Edward Wilson-Ewing for comments on an early draft.   

\bibliographystyle{apsrev}
\bibliography{LQCAlpha}

\end{document}